\documentclass[useAMS,usenatbib]{mnras}

\usepackage{graphicx}
\usepackage{amssymb}
\usepackage{array}

\usepackage{ulem}
\usepackage[dvipsnames]{xcolor}

\newcommand{\kms}{km\,s$^{-1}$}
\newcommand{\HI}{\textrm{H}\,\textsc{i}}

\newcommand{\Msun}{\mbox{$\mathcal{M}_{\sun}$}}
\newcommand{\Mdisc}{\mathcal{M}_{\rm disc}}

\newcommand{\Mtotal}{\mathcal{M}_{\rm total}}
\newcommand{\Vh}{V_{\rm h}}
\newcommand{\VLG}{V_{\rm LG}}
\newcommand{\VCMB}{V_{\rm 3K}}
\newcommand{\Vrot}{\mbox{$V_{\rm rot}$}}
\newcommand{\Vmax}{\mbox{$V_{\rm max}$}}
\newcommand{\Wm}{\mbox{$W^c_{m50}$}}

\newcommand{\Nancay}{Nan\c{c}ay}

\graphicspath{ {./figs/} }

\newcommand{\Ntotal}{397}
\newcommand{\Nsample}{331}

\title[The TF relation for flat galaxies]{The Tully-Fisher relation for flat galaxies}

\author[D.\ I.\ Makarov, N.\ A. Zaitseva \& D.\ V.\ Bizyaev]
{
D.\ I.\ Makarov,$^1$\thanks{E-mail: dim@sao.ru (DIM)}
N.\ A.\ Zaitseva,$^2$
D.\ V.\ Bizyaev$^{3,2}$
\\
$^1$ Special Astrophysical Observatory of RAS, Nizhnij Arkhyz, Karachai-Circassia 369167, Russia
\\
$^2$ Sternberg Astronomical Institute, Moscow State University, Universitetskij pr.\ 13, 119991 Moscow, Russia
\\
$^3$ Apache Point Observatory and New Mexico State University, Sunspot, NM 88349, USA
}

\begin{document}

\label{firstpage}

\pagerange{\pageref{firstpage}--\pageref{lastpage}}


\maketitle

\begin{abstract}
We construct a multiparametric Tully-Fisher (TF) relation for a large sample of edge-on galaxies from the Revised Flat Galaxy Catalog
using \HI{} data from the EDD database and parameters from the EGIS catalog.
We incorporate a variety of additional parameters including structural parameters of edge-on galaxies in different bandpasses.
Besides the rotation curve maximum, only the \HI{}-to-optical luminosity ratio and optical colours  
play a statistically significant role in the multiparametric TF relation.
We are able to decrease the standard deviation of the multiparametric TF relation down to 0.32 mag,
which is at the level of best modern samples of galaxies used for 
studies of the matter motion in the Universe via the TF-relation.
\end{abstract}


\section{Introduction}
\label{sec:intro}

\citet{Karachentsev1989} notes that highly inclined, or "flat" galaxies are
a suitable tool to investigate bulk motion of galaxies on the scales up to 200~Mpc.
Such objects have a number of advantages for observations.
Galaxies with apparent axes ratio $a/b\ge7$ from \citet{Karachentsev1989}  belong to the late morphological types Sc--Sd.
Since the flat galaxies are observed almost edge-on, they have a high surface brightness, 
making it easier to detect and classify them even at large distances.
Edge-on galaxies do not require a correction for their inclination, 
which usually introduces the most of uncertainties into the Tully-Fisher (TF) relation.
Besides that, the Sc--Sd galaxies show more uniform spatial distribution in the local Universe than galaxies of early types. 
The principal disadvantage of using the flat galaxies is the complexity of accounting for their internal extinction.
Highly inclined galaxies are routinely excluded from the TF distance estimations 
in order to avoid unpredictable effects of strong light absorption.
Nevertheless, the very first works by \citep{Karachentsev1989,Karachentsev1991}
show that the linear diameter and luminosity of flat galaxies are good distance indicators for the TF relation.

The Flat Galaxies Catalogue \citep[FGC;][]{FGC} and its revised version \citep[RFGC;][]{RFGC} 
have been designed for investigations of large-scale bulk motion of galaxies on the scales of 100--200~Mpc.
RFGC contains 4236 galaxies all over the sky, with blue axes ratio $a/b\ge7$
and maximal diameter $a\ge0\farcm6$.
The peculiar velocities of the flat galaxies have been studied by
\citep{KKKP1995,KKKP2000,KKKMP2000,PKKK2001,KKK+2003,PT2004,Kashibadze2008,PP2010}
using multiparametric approach to the TF relation.
Most of those studies are based on the linear diameter of galaxies as a distance indicator.
They obtain a typical scatter of 0.6--0.7~mag for the TF relation for RFGC galaxies,
depending on the choice of the optical or near-infrared passbands \citep{KKK1997,KMK+2002,Kashibadze2008}.

Significant progress in quantity, quality and homogeneity of observing data for the RFGC galaxies has been achieved
in the frames of the `Cosmic Flows' program: a large sample of edge-on galaxies has been observed
with the Green Bank 100-m and the Parkes 64-m telescopes \citep{CTM+2011}.
These observations in combination with archival radio data from other telescopes have been processed
in a uniform manner in order to obtain a homogeneous set of \HI{}-linewidth
with precision better than 20~\kms{} \citep{EDD:HI}.
At the same time, the catalogue of edge-on disc galaxies \citep[EGIS;][]{EGIS} based on the 
Sloan Digital Sky Survey data \citep{SDSS:DR7} provides
parameters of aperture photometry in the optical $g$, $r$ and $i$ bands as well as their structural parameters
for 5747 genuine edge-on galaxies.
The combination of high precision radio and optical data gives us an opportunity to study
the luminosity-linewidth relation for edge-on galaxies with high accuracy and to incorporate additional 
parameters in attempt to minimize the TF relation scatter.

Note that in the case of true edge-on galaxies we can directly study the vertical distribution of luminous matter. 
This may matter for the TF relation since \citet{ZMM1991} show that 
the relative thickness of a gravitationally stable galactic disc depends on relative mass of a spherical component
and traces the contribution of the dark matter into the galactic mass
\citep{ZBM2002,BizyaevMitronova2002,KregelVanDerKruit2005,SR2006}.
The existence of superthin galaxies with $a/b\ge10$ is possible only in the presence of massive dark halos around galaxies \citep{Bizyaev+2017}.
Using N-body simulations, \citet{ZMM1991,SR2006,Khoperskov+2010} show that theoretical relationship
between the thickness of disc galaxies and the mass of their spherical components
allows one to estimate the lower limit for the dark halo mass in galaxies. 
The latter may affect the TF relation, which encourages us to consider the relative thickness of galactic discs as an additional parameter in the TF relation.

\section{The sample, Data and Corrections}
\label{sec:sample}

\begin{figure}
\centerline{
  \includegraphics[width=0.46\textwidth,clip]{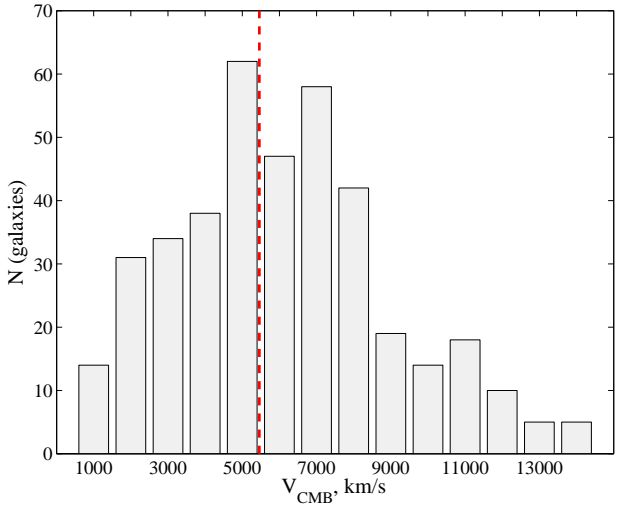}
}
\caption{
Distribution of RFGC galaxies with reliable photometry from the EGIS catalogue and \HI{}-data from EDD.
The vertical dashed line indicates the effective depth of the sample.
}
\label{f:cz3k}
\end{figure}

We are able to select \Ntotal{} flat galaxies from the RFGC catalogue \citep{RFGC} 
with precise measurement of \HI{} linewidth (error $<20$~\kms{}) 
using the All Digital \HI{} Profile Catalog \citep{EDD:HI} form the Extragalactic Distance Database \citep[EDD;][]{EDD} 
and photometry from the EGIS catalogue \citep{EGIS}. 
The distribution of selected galaxies is shown in Fig.~\ref{f:cz3k} in the cosmic microwave background (CMB) reference frame.
The effective depth of the sample is 5459~\kms{}. 

We excluded 14 galaxies with radial velocity $\VLG\le1000$~\kms{} from the consideration
to avoid biased distance estimation from the Hubble law on small redshifts.
For the same reason, we eliminated 15 objects which reside within the `zero-velocity surface' of Virgo cluster
with the distance to the cluster centre less than 7.2~Mpc \citep{KTW2014}.
At the distance of 16.5~Mpc \citep{MBC+2007} it corresponds to the angular radius of 23.6 deg
and spans from $-321$ to $2679$~\kms{} with respect to the Local Group.
Finally, we rejected 12 more objects with high Galactic extinction, $A_B\ge0.6$.

At a glance, only a few objects show large deviation from the general TF trend.
We found that PGC\,15031 and PGC\,91372 should be excluded from consideration because of problems with their \HI{}-data.
Moreover, we carried out visual inspection of all galaxies from the sample to exclude objects
with suspicious photometry due to relatively bright stars superimposed on body of the galaxy, 
with signs of interaction, and with nearby galaxies that may contaminate the radio fluxes.
We provide comments on 14 removed galaxies in Appendix \ref{appendix:listX}.
As a result, our list of good RFGC-galaxies consists of \Nsample{} objects.

The All Digital \HI{} Profile Catalog \citep{EDD:HI} of EDD provides homogeneous information on \HI{}-line measurements
from original observations carried out in framework of the NRAO Cosmic Flows Large Program
as well as archival data form the Arecibo, Parkes, Green Bank, and \Nancay{} telescopes.
In this paper we use the line width at the 50 per cent of the mean flux level within the \HI{} signal, or $W^c_{m50}$ form EDD.
This value is already corrected for a slight relativistic broadening 
and the broadening caused by finite spectral resolution as described by \citet{CT2015}.
We do not apply a correction for the inclination because of most of our objects have $i>86^\circ$~\citep{EGIS} and the correction is negligible.

We incorporate the structural parameters of the galaxies determined by EGIS~\citep{EGIS}:
the radial and vertical scales, central surface brightness and bulge-to-disc ratio.
The total luminosity was estimated in three SDSS passbands: $g$, $r$ and $i$, 
within encompassing ellipses at the level of signal-to-noise $S/N=2$ pixel$^{-1}$.
All magnitudes were corrected for Galactic foreground extinction using maps by \citet{DustMap98} as described by \citet{EGIS}.
In the present paper we also apply the K-correction to the magnitudes
according to the methodology\footnote{\url{http://kcor.sai.msu.ru/}} described by \citet{Kcorr10} and by \citet{Kcorr12}.
In the case of SDSS $g$-band we average the K-corrections calculated for two cases: $(g-r)$ and $(g-i)$ colours.
For comparison with literature data, we transformed the SDSS-magnitudes to the Johnson-Cousins $B$ and $I_c$,
using equations in SDSS algorithms\footnote{\url{http://www.sdss.org/dr12/algorithms/sdssubvritransform/}}.
These data are compared with results by \citet{Karachentsev+2017} and \citet{IbandTF}.

At the first step we do not apply any correction for the internal extinction.
Note that there are some reasons to neglect
the internal extinction effects in the multiparametric approach. 
Typically, the dust is concentrated in a narrow layer in the plane of the galactic disc.
Hence, in edge-on galaxies significant part of light is not affected by dust.
Moreover, the RFGC-galaxies form narrow distribution by morphology
with a predominance of late types, Sc--Sd \citep{UFG1}.
We expect that internal extinction is comparable in similar objects
and its influence varies smoothly with the change of galaxy properties 
such as total mass, stellar populations or relative thickness of the disc.
In the case of the multiparametric TF relation we expect that the internal extinction
will be partially taken into account through considered additional parameters,
such as colour, relative mass of hydrogen, relative thickness and amplitude of rotation curve.

As follows from the Fig.~\ref{f:cz3k}, the effective depth of our sample is 75~Mpc.
On the scales up to 100--200~Mpc the peculiar velocity field in the Universe demonstrates complicated structure \citep[see, for instance,][]{LaniakeaSC}.
The bulk motion of galaxies on the 100~Mpc scale is about 270~\kms{} with respect to the CMB rest frame \citep{Hoffman+2015}.
The large scale flows of galaxies and peculiar motion of the Milky Way complicate the distance estimation from the Hubble law,
hence increasing the TF relation dispersion.
Motivated by this warning, we check the scatter of the TF relation in dependence of the reference frame choice:
the CMB, the Local Group centroid or the bulk motion of the sample itself.
In the case of predefined Solar apexes as CMB \citep{apexCMB} and the Local Group centroid \citep{apexLG},
we calculate the dispersion of the TF relation using a simple least-square minimization,
where the observational heliocentric radial velocity was corrected for the corresponding Solar motion.
In the case of the bulk motion we use a non linear least-square minimization of the TF relation (the zero point and the slope are the model parameters) 
with simultaneous variation of three parameters of the Solar apex.
Because the SDSS survey covers only a part of the sky, 
we expect that the quality of the fit will be much better in the direction of the Galactic North Pole than perpendicular to it.
Thus, we also check a model with only one free parameter of the Solar apex in the direction on the Galactic North Pole.
Our analysis shows that the TF relation standard deviation lies in the range from 0.34 to 0.41~mag for considered Solar apexes.
The correction to CMB apex gives the minimal scatter among all possibilities in all passbands: 0.40, 0.35 and 0.34~mag in the 
$g$-, $r$- and $i$, respectively.
Thus, we chose the CMB rest frame as the best one for consequent analysis.
Interesting to note that the scatter is significantly larger in the $g$-band than in other filters, 
which we refer to the effects of internal extinction in edge-on galaxies.

Throughout this paper we use cosmology parameters $\Omega_{\Lambda}=0.7$, $\Omega_{m}=0.3$ and $H_{0}=73$~\kms{}.

\section{The Tully-Fisher Relation}
\label{sec:tf}

\begin{figure}
\centerline{\includegraphics[width=0.47\textwidth,clip]{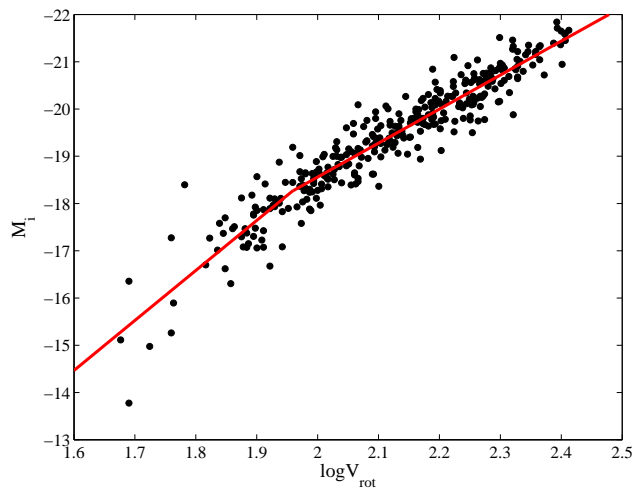}}
\caption{
The TF relation for selected RFGC-galaxies in the $i$-band. 
The 
trends
in the $g$ and $r$ bands are similar.
}
\label{f:TF}
\end{figure}

Figure~\ref{f:TF} shows the TF relation between the absolute magnitude in the $i$-band and the rotational velocity $\Vrot=\Wm/2$.
The relation looks similar in the $g$ and $r$ bands.
The absolute magnitudes are not corrected for the internal extinction.
It is seen that the fast rotating galaxies form a tight linear relation,
while dwarf galaxies with $\Vrot\lesssim91$~\kms{} lie systematically below the relation extrapolated from giants.
To address this difference, we fit the data with a broken line:
\begin{equation}
M = \left\{
\begin{array}{ll}
a + b \log\Vrot       & \textrm{if } \Vrot > V_{\rm bp} \\
\hat{a} + c \log\Vrot & \textrm{if } \Vrot \le V_{\rm bp} \\
                      & \hat{a}=a+(b-c)\log V_{\rm bp},
\end{array}
\right.
\label{e:brokenLine}
\end{equation}
\noindent 
where $a$ and $b$ refer to the intercept and slope terms for the giant galaxies, respectively;
$V_{\rm bp}$ is the break point of the relation;
$c$ is the slope term for the dwarf galaxies;
and $\sigma$s represent the scatter for the giant and whole-sample relation, respectively.
The results are summarized in Table~\ref{t:TF}.

\begin{table}
\caption{
The Broken Line fitting of the TF relation in $g$, $r$ and $i$ bands.
The terms $a$, $b$, $V_{\rm bp}$ and $c$ are defined in the equation~\ref{e:brokenLine}, and
correspond to the intercept and slope for the giants, 
the rotation speed for transition between giants and dwarfs, and 
the slope for TF relation for dwarfs, correspondingly.
}
\label{t:TF}
\begin{tabular}{cr@{ $\pm$ }lr@{ $\pm$ }lr@{ $\pm$ }l}
\hline\hline{}
                &\multicolumn{2}{c}{$g$}     & \multicolumn{2}{c}{$r$}    & \multicolumn{2}{c}{$i$}    \\
\hline
$a$             & $-6.18$ & 0.10             & $-5.07$   & 0.07           & $-4.18$   & 0.04           \\
$b$             & $-5.85$ & 0.10             & $-6.63$   & 0.10           & $-7.19$   & 0.06           \\
$V_{\rm bp}$    &  91.0   & 1.5              & 91.0      & 1.3            & 91.0      & 0.8            \\
$c$             & $-9.57$ & 0.25             & $-10.09$  & 0.21           & $-10.58$  & 0.19           \\
\hline
$\sigma$ (giants) & \multicolumn{2}{c}{$0.40$} & \multicolumn{2}{c}{$0.35$} & \multicolumn{2}{c}{$0.34$} \\
$\sigma$ (total)  & \multicolumn{2}{c}{$0.44$} & \multicolumn{2}{c}{$0.41$} & \multicolumn{2}{c}{$0.40$} \\
\hline\hline
\end{tabular}
\end{table}

Note that similar TF break was reported by \citet{BaryonicTF}
using a sample of galaxies with circular velocities ranging between  $30\lesssim V_{c} and \lesssim300$~\kms{}. 
Field dwarfs with $V_{c}\lesssim90$~\kms{} fall bellow the relation defined by more rapidly spinning galaxies.
\citet{BaryonicTF} noted that that these faint galaxies are very gas rich
and the bulk of their baryonic material is still not converted into stars.
Thus, these slowly rotating galaxies are underluminous with respect to the bright galaxies and to the extrapolation of the TF
relation inferred from them.
Using the sum of the stellar and gas masses restores the linear relation over the entire observed range \citep{BaryonicTF}.
Authors argue that the traditional TF relation is a particular case of more fundamental Baryonic Tully-Fisher (BTF) relation 
between the total mass of baryons and rotational velocity.

This explanation is also suitable for our subsample of thin and slowly rotating, gas-rich galaxies.
As it will be shown in section~\ref{sec:TFplus}, 
the multiparametric TF relation restores a linear behaviour 
when the \HI{} mass and colours of stellar populations are incorporated into the analysis.
At the same time we can not exclude the influence of selection biases.
On average, the discs of dwarf galaxies are thicker in the comparison to regular spiral galaxies.
Thus, \citet{Disc_thickness} found that below $\mathrm{M}^{*}\approx2\times10^{9}$~\Msun{} low-mass galaxies become systematically thicker.
This luminosity corresponds to $M_{i}\sim-18$~mag, which coincides with the absolute magnitude of the break for the galaxies in our sample (see Fig.~\ref{f:TF}).
The strict cut of the axes ratio of $a/b\geq7$ could create the preference of mostly thin, low surface brightness galaxies in our sample. 
These galaxies have lower luminosity comparable to the whole sample of dwarf galaxies.

The TF relation with a break was also identified in simulations.
\citet{Guo+2011} applied a semi-analytic model of the galaxy formation to the Millennium \citep{Millenium} and Millennium-II simulations \citep{MilleniumII}.
Their predicted TF relation shows the break of the linear relation near $\log \Vmax \simeq 2.0$ \citep[][see Fig. 13]{Guo+2011}.
This value corresponds to $M_{r}-5\log(h)\simeq-18$, which it is good agreement with our TF relation for the edge-on galaxies.

\section{Impact of the dust}
\label{sec:dust}

\begin{figure}
  \centerline{\includegraphics[width=0.47\textwidth,clip]{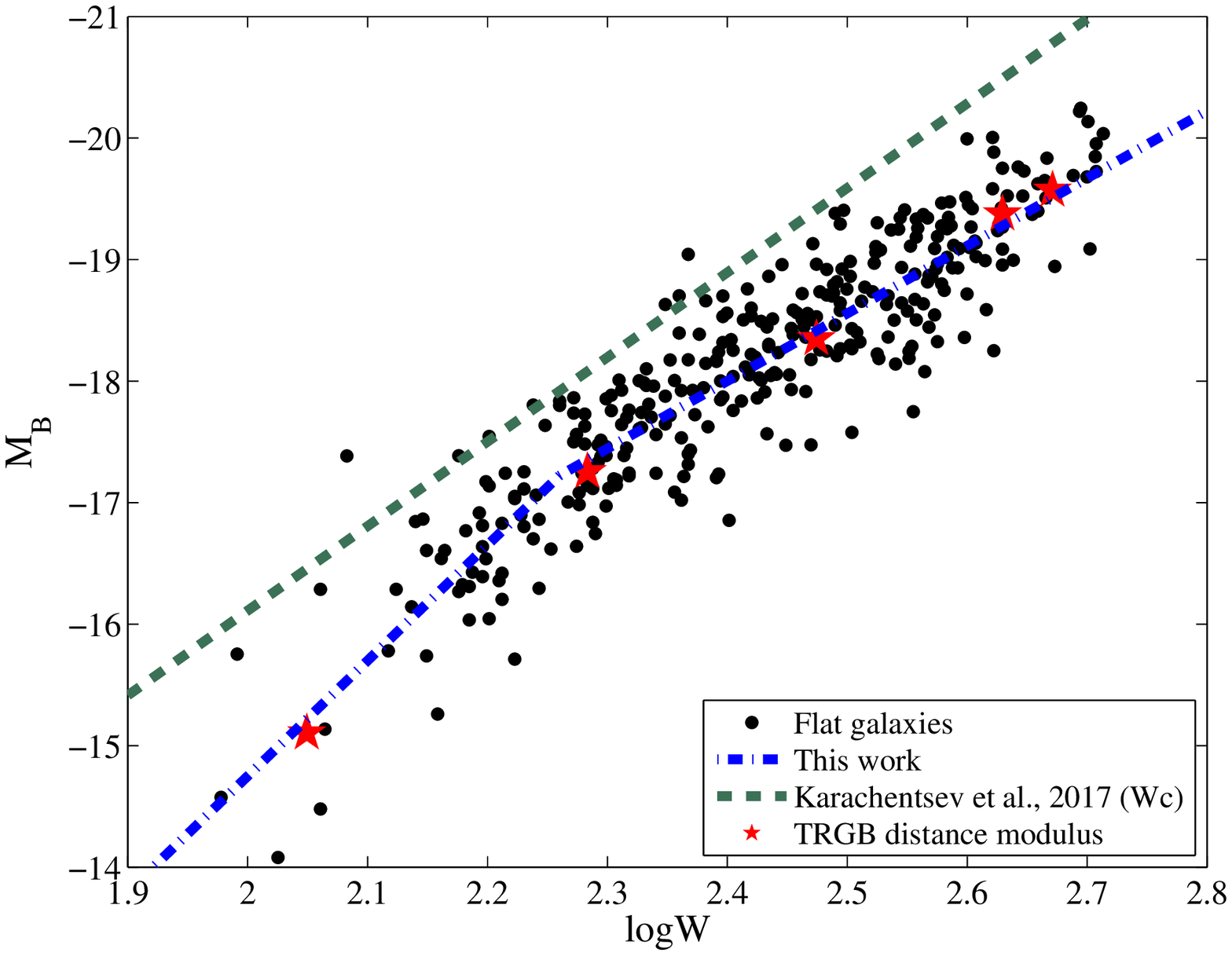}}
  \centerline{\includegraphics[width=0.47\textwidth,clip]{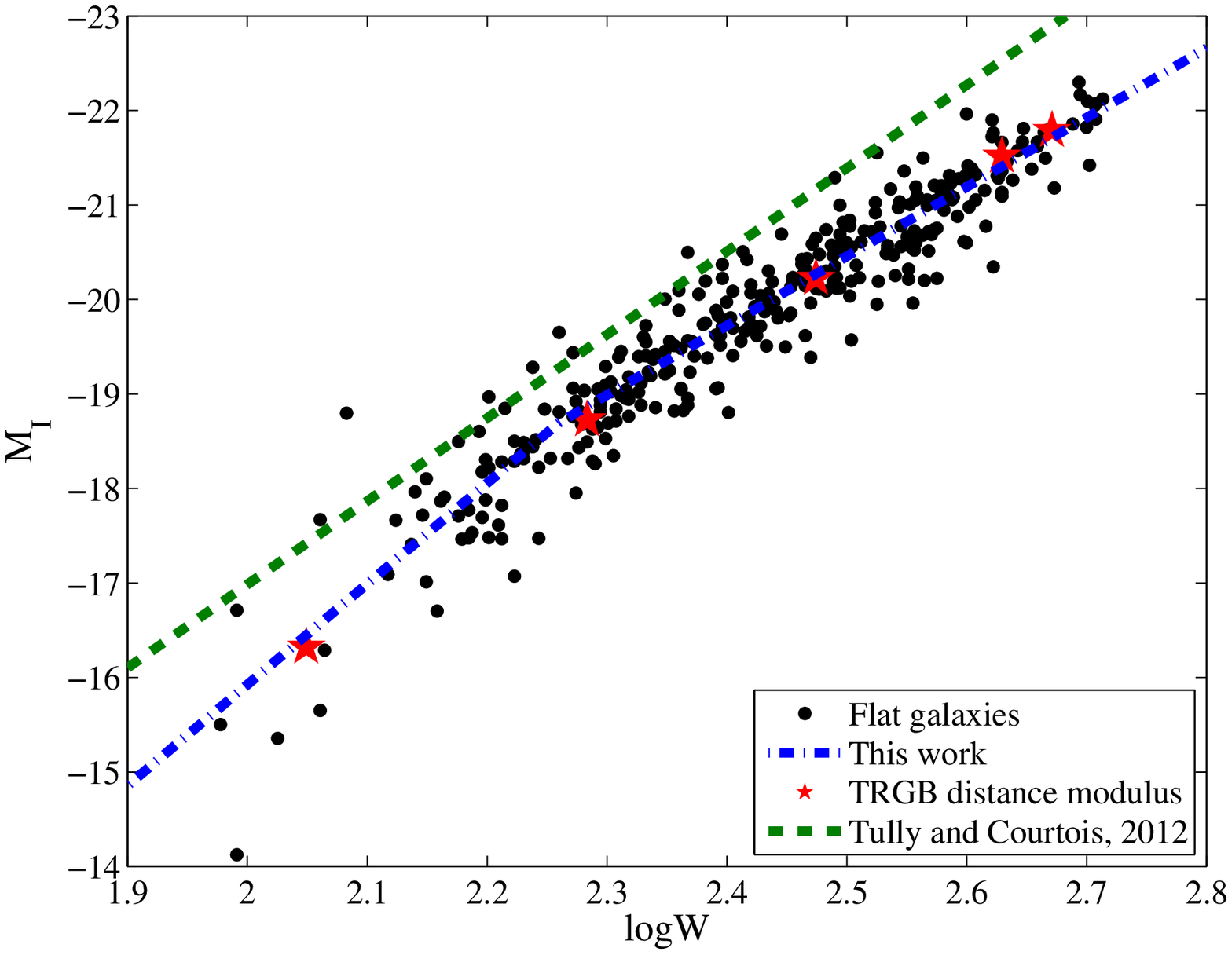}}
\caption{
The absolute magnitude -- line-width relation for RFGC-galaxies (black dots) in the $B$ (the top panel) and $I_{C}$ bands (the bottom panel).
The galaxies with known TRGB distances are marked by the red stars.
Our broken line approximation is shown by the dash-dotted blue line.
The green dashed line corresponds to the $B$-band TF relation for the Local Volume galaxies \citep{Karachentsev+2017} in the top panel,
and to $I_{C}$-band relation for the template set of giants from the Cosmicflows-2 project \citep{IbandTF} in the bottom panel.
}
\label{f:TF-compare}
\end{figure}

As noted by \citet{Dust120kms}, the dust distribution in edge-on galaxies depends on their rotation velocity.
Fast spinning galaxies ($\Vrot>120$~\kms{}) show well-defined dust lines, 
while slow rotating galaxies do not have a regular dust line.
The distribution of the dust in low-mass galaxies with $\Vrot<120$~\kms{} has a large scale height 
and appears more diffuse in the comparison with massive objects.
\citet{Bizyaev+2017} also notice that the gas and dust layer thickness is comparable to the stellar disc thickness
in small ($\Vrot \lesssim 100$~\kms{}), blue and very thin disc galaxies. 
Note that the 120~\kms{} threshold is not so far from the break of our TF relation shown in Fig.~\ref{f:TF}.
\citet{Dust120kms} demonstrate that all high-mass galaxies with dust lanes are gravitationally unstable.
The change in the slope of TF relation may reflect the dependence of the disc stability, dust structure and star formation efficiency on the mass of galaxies.

The interconnection between the mass and dust content in galaxies makes the apparent slope of the TF relation less steep.
Variety of the dust distribution patterns adds up an additional scatter to the relation for high-mass galaxies.
In the Fig.~\ref{f:TF-compare} we compare the TF relations for our sample of the edge-on galaxies with corresponding relations
for the nearby Local Volume galaxies \citep{Karachentsev+2017}
and well calibrated set of giants used in the Cosmicflows-2 project \citep{IbandTF}.
The edge-on galaxies not corrected for the internal extinction appear systematically less luminous 
in the comparison to the normal galaxies in which the correction for the internal extinction was applied.
We can statistically estimate the global internal extinction in the giant edge-on galaxies via subtracting one TF relation from the other.
It gives $A_{B}= 1.39 \log W-2.45$ in the $B$-band and $A_{I}= 1.46 \log W-2.76$ in the $I_{C}$-band.

\section{Multiparametric TF relation}
\label{sec:TFplus}

We attempt to improve the TF relation by including various distance independent values in the regression.
As such parameters, we consider the EGIS structural parameters in the $g$, $r$, $i$ bands,
the sizes of the galaxies in the blue and red bands reported by \citep{RFGC},
and parameters from the All Digital \HI{} Profile Catalog \citep{EDD:HI} 
and the HyperLEDA database\footnote{\url{http://leda.univ-lyon1.fr}} \citep{LEDA}.
The list of tested parameters also includes 
the minor-to-major axes ratio, $log(\frac {b}{a})_{o,e}$ in the blue and red POSS-I bands;
the vertical-to-radial scale ratio $\log(\frac {z}{h})_{g,r,i}$, in the $g$, $r$ and $i$ bands;
the \HI{} colour index, $m_{21}-\{g,r,i\}$, which corresponds to the \HI-mass-to-light ratio, 
where $\{g,r,i\}$ is one of a galaxy visible magnitude in the $g$, $r$, and $i$,
and $m_{21}$ is the 21-cm line flux expressed in magnitude according to $m_{21}=-2.5\log F+17.40$ \citep{RC3};
galactic colours in different bands using aperture photometry of SDSS images.
Also we checked and ensured that there is no dependence on the position of galaxies in the sky or the Galactic extinction in the B-band.
In addition, we test if the ratios $\log\Vrot\log(\frac{b}{a})_{o,e}$ and $\log\Vrot\log(\frac{z}{h})_{g,r,i}$
are connected with the internal extinction in our edge-on galaxies.

We use stepwise regression to select significant parameters of the multiparametric TF relation.
This iterative procedure adds and removes terms from multilinear model based on their statistical significance in a regression.
At each step, the $p$-value of the F-statistics is computed to test the models with and without a potential term.
The term is added to corresponding model if there is an evidence that the term coefficient would significantly differ from zero.
Conversely, the term is removed from the model if there is insufficient evidence that the coefficient differs from a zero.
The iterations terminate when no single step improves the model.
We use MATLAB's realisation of this algorithm with $p$-value of 0.001 as threshold for the including/excluding the parameters.

\begin{figure}
\centerline{\includegraphics[width=0.47\textwidth,clip]{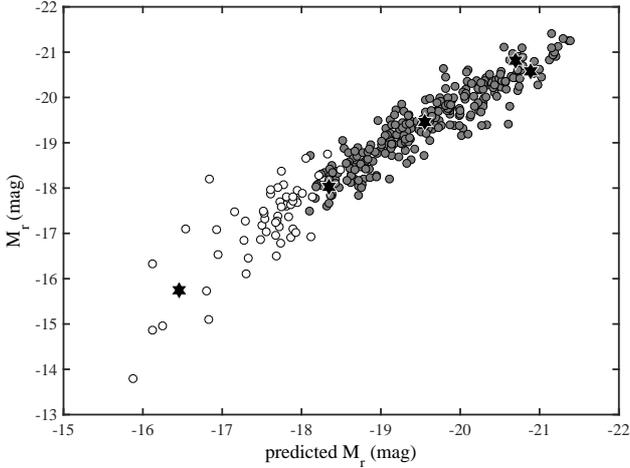}}
\caption{
The multiparametric TF relation for RFGC-galaxies in the $r$-band. 
The relations in the $g$ and $i$ bands look similar.
The regression is obtained for massive galaxies (filled circles) with $\log\Vrot>1.96$.
The open circles show the extrapolation to the slow rotating galaxies.
The objects with known TRGB-distances are designated by filled hexagons.
}
\label{f:mpTF-big}
\end{figure}

\begin{table}
\caption{
Regression coefficients for the multiparametric TF relation in different SDSS bands. 
Subsample of the massive galaxies contains the objects with $\log\Vrot>1.96$.{}
}
\label{t:mpTF-big}
\begin{tabular}{crrr}
\hline\hline
& \multicolumn{1}{c}{$M_g$} & \multicolumn{1}{c}{$M_r$} & \multicolumn{1}{c}{$M_i$} \\
\hline
$\log\Vrot$                & $-8.60\pm 0.32$ & $-8.22\pm 0.28$ & $-8.48\pm 0.30 $ \\
$(m_{21}-\{g,r,i\})$       & $-0.23\pm 0.04$ & $-0.23\pm 0.04$ & $-0.22\pm 0.04 $ \\
$(g-i)$                    & $ 2.00\pm 0.19$ & $             $ & $ 1.19\pm 0.20 $ \\
$(g-r)$                    & $             $ & $ 2.21\pm 0.28$ & $              $ \\
zero-point                 & $-2.09\pm 0.10$ & $-2.80\pm 0.09$ & $-2.28\pm 0.10 $ \\
\hline
$\sigma$                   & 0.34            & 0.32            & 0.32            \\
\hline\hline
\end{tabular}
\end{table}

The regression coefficients for the sample of 276 massive, fast rotating galaxies with $\log\Vrot>1.96$ are presented in Table~\ref{t:mpTF-big}.
The last line shows the resulting scatter of the relation.
The zero-point was calibrated through the galaxies with known, redshift-independent distances based on 
the tip of the red giant branch (TRGB; see Sect.~\ref{sec:zero-poit}).
In addition to the rotational velocity, the most significant parameters are the optical and \HI{} colour indexes.
It reflects the fact that the stellar population and gas play significant role in the TF relation.
These parameters have high confidence level with the $p$-value less than $10^{-8}$.
No more considered parameters were included in the final relation because their $p$-value is greater than 0.04.
Surprisingly, the multiparametric relation for giants improves the behaviour of the whole sample including 55 slow rotators with $\log\Vrot<1.96$.
Taking into account the optical and \HI{} colour indexes for giant galaxies removes the break and suppresses the dispersion for the dwarfs also.
The comparison between the predicted and observed absolute $r$-magnitudes of galaxies is shown in Fig.~\ref{f:mpTF-big}.

\begin{figure}
\centerline{\includegraphics[width=0.47\textwidth,clip]{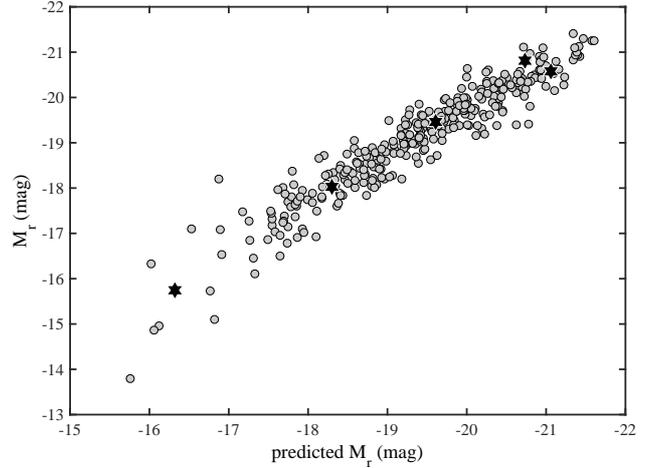}}
\caption{
The multiparametric TF relation for the whole sample of 331 flat galaxies in the $r$-band. 
The hexagons show the galaxies with known, TRGB-distances.
}
\label{f:mpTF-all}
\end{figure}

\begin{table}
\caption{
Regression coefficients for multiparametric TF relation in different SDSS bands. 
}
\label{t:mpTF-all}
\begin{tabular}{crrr}
\hline\hline
& $M_g$ & $M_r$ & $M_i$ \\
\hline
$\log\Vrot$                & $-8.50\pm 0.26$ & $-8.48\pm 0.25$ & $-8.57\pm 0.28$ \\
$(m_{21}-\{g,r,i\})$       & $-0.30\pm 0.04$ & $-0.30\pm 0.04$ & $-0.29\pm 0.04$ \\
$(g-i)$                    & $             $ & $             $ & $ 1.12\pm 0.21$ \\
$(g-r)$                    & $ 2.88\pm 0.29$ & $ 2.24\pm 0.30$ & $             $ \\
zero-point                 & $-2.21\pm 0.09$ & $-2.29\pm 0.09$ & $-2.05\pm 0.11 $ \\
\hline
$\sigma$ (giants)          & 0.34            & 0.32            & 0.32            \\
$\sigma$                   & 0.40            & 0.38            & 0.39            \\
\hline\hline
\end{tabular}
\end{table}

The Fig.~\ref{f:mpTF-big} suggests that a simple, linear multiparametric regression can be constructed for the full sample of 331 edge-on galaxies.
The resulting relation is shown in Fig.~\ref{f:mpTF-all} and the corresponding coefficients are reported in Table~\ref{t:mpTF-all}.
Similar to the sample of giants-only, the most significant terms are the optical and \HI{} colour indexes, 
which show the $p$-value less than $10^{-6}$.
The $(r-i)$ term appears at the level of $p\simeq0.004$ for the relation in the $i$-band.
The ratio of the vertical scale heights in different bands  $log\frac{z_r}{z_g}$ and $log\frac{z_i}{z_g}$ contributes with 
the $p$-value of 0.04 and 0.02, respectively.
Replacement of the giant-only sample by the whole sample does not affect the small scatter of the 
giant galaxies with respect to the whole-sample regression line. 
We come to a conclusion that applying the multiparametric linear regression in lieu of the broken-line parameterisation 
improves the standard deviation of the relation by 10 per cent.

\section{Zero-point calibration}
\label{sec:zero-poit}

\begin{table}
\caption{
List of the TRGB distance measurements.
}
\label{t:TRGB}
\begin{tabular}{rrrll}
\hline\hline
\multicolumn{1}{c}{PGC}  & \multicolumn{1}{c}{RFGC} & \multicolumn{1}{c}{EGIS}             & \multicolumn{1}{c}{$(m-M)_0$}    &       \\
\hline
 6699 &  384 & EON 027.382 +32.589 & $28.61\pm0.01$ & $^a$ \\
      &      &                     & $28.53$        & $^e$ \\
      &      &                     & $28.55\pm0.03$ & $^g$ \\[3pt]

24930 &      & EON 133.173 +33.421 & $29.96\pm0.38$ & $^a$ \\
      &      &                     & $29.86\pm0.06$ & $^b$ \\[3pt]

39422 & 2245 & EON 184.374 +37.807 & $28.17\pm0.07$ & $^a$ \\ 
      &      &                     & $28.21\pm0.11$ & $^f$ \\
      &      &                     & $28.19\pm0.12$ & $^d$ \\
      &      &                     & $28.20\pm0.03$ & $^h$ \\
      &      &                     & $28.26\pm0.24$ & $^k$ \\
      &      &                     & $27.88\pm0.17$ & $^i$ \\
      &      &                     & $28.16\pm0.08$ & $^j$ \\[3pt]

41618 & 2315 & EON 188.190 +00.115 & $29.61\pm0.12$ & $^a$ \\
      &      &                     & $29.61\pm0.21$ & $^c$ \\[3pt]

54470 & 2946 & EON 228.973 +56.329 & $31.19\pm0.05$ & $^a$ \\
      &      &                     & $31.13\pm0.10$ & $^f$ \\
\hline\hline
\end{tabular}
\begin{tabular}{p{0.4\textwidth}}
$^a$ EDD CMD/TRGB 2016/04/27 \citep{EDD:CMD} \\
$^b$ \citet{KTM2015} \\
$^c$ \citet{KTW2014} \\
$^d$ \citet{MMU2013} \\
$^e$ \citet{TG2012} \\
$^f$ \citet{RSdJS2011} \\
$^g$ \citet{TRD2006} \\
$^h$ \citet{SDdJ2005} \\
$^i$ \citet{MFR2005} \\
$^j$ \citet{TG2005} \\
$^k$ \citet{KSD2003} \\
\end{tabular}
\end{table}

Because of strict limit on angle of view, the edge-on galaxies are quite rare.
There is only small number of nearby FGC galaxies available for distance determination with precise methods.
There are no distance measurements from Cepheids for the RFGC galaxies.
In total, only eleven RFGC-galaxies have TRGB-distance measurements,
and only five of them are included in our sample.
Most of them have distance less than 10~Mpc, which is an effective distance for the TRGB measurements with the Hubble Space Telescope.
We gathered these distance in Table~\ref{t:TRGB},
which contains the object ids by the HyperLeda database, 
RFGC and EGIS catalogues,
together with the TRGB distance moduli from the literature.
Note that these distances were estimates using different TRGB-techniques and calibrations.
EDD provides the most homogeneous and precise set of the data.
Most important, the data are available for all five galaxies of our sample.
Thus, we decide to use EDD distances for the zero-point calibration.
It should be noted that because of the small distances, these galaxies were not included in the regression analysis, 
but we use them for zero-point calibration of the final multiparametric TF relation.
The result is shown in corresponding line in Tables~\ref{t:mpTF-big} and \ref{t:mpTF-all}.

\section{Discussion and conclusions}
\label{sec:conclusion}

We construct a multiparametric TF relation for a sample of 331 edge-on RFGC galaxies.
These objects have precise \HI{} linewidth measurements with uncertainty better than 20~\kms{}
from the All Digital \HI{} Profile Catalog \citep{EDD:HI} of the EDD database \citep{EDD}.
The optical photometry and structural parameters of the galaxies were taken from the SDSS-based EGIS catalogue. 

The classical TF relation between the absolute magnitude and the \HI{} linewidth considered for our sample of edge-on galaxies
reveals different trends for the giant and dwarf galaxies.
The dwarfs with the rotation curve amplitude $\Vrot<91$~\kms{} are under-luminous with respect to the 
extrapolation of the giant-based TF relation.
This fact can be well explained in the framework of baryonic TF paradigm.
Small gas reach dwarf galaxies hold sufficient fraction of baryonic matter in their gas still not converted into stars.
Thus, their optical luminosity is lower than expected for the giant galaxies with high star formation efficiency.
Using the total mass of baryons instead of the stellar luminosity takes this effect into account
and puts the giant and dwarf galaxies on the same relation \citep{BaryonicTF}.
We show  that both slowly and highly rotating galaxies obey the same relation on
our multiparametric TF relation made for for edge-on galaxies.

Our searches for additional significant parameters show that the multiparametric relation can expand the classical TF law with two terms:
the \HI{} colour index and optical colors of the galaxies. 
The \HI{} colour index $m_{21}-\{g,r,i\}$ reflects the fraction of \HI{} mass with respect to the optical luminosity.
The colours $(g-i)$ or $(g-r)$ indicate properties of stellar population, such as mean age and metallicity.
We find that adding any other structural parameters or their combinations make statistically negligible effect on the TF relation.
Our analysis shows that only parameters directly related to the baryonic matter, 
i.e. stellar population and fraction of gas, play significant role in the TF relation.
Note that the coefficients of the multiparametric TF relation are almost independent of the passband.
This differs from the simple \HI{}-linewidth--absolute magnuitude relation where the slope significantly changes with the passband used, which can be seen from Table~\ref{t:TF}.
Thus, taking into account the optical and \HI{} colour indexes helps us find a universal relation.

The zero-point of our multiparametric TF relation was calibrated via 5 nearby edge-on galaxies with precise distance measurements based on the TRGB method.

The insignificant influence of the relative disc thickness on the TF relation requires an additional discussion.
It is known that the presence of a spheroidal component stabilizes galactic discs.
Therefore a disc submerged in a halo should have lower values of the vertical velocity dispersion, 
and should appear more thin.
\citet{ZMM1991} show that the minimal relative thinkness of a collisionless disc decreases with its relative mass, 
$z/h\sim \Mdisc/\Mtotal$ \citep{ZBM2002}, 
where $\Mdisc$ and $\Mtotal$ are masses of the disc component and total mass, respectively.
One can expect that this connection have to be reflected by the TF relation for the case of edge-on galaxies.
However, we find that the stellar disc relative thickness have no significant effect on the multiparametric TF relation.

We see several possible explanations.
The stellar parameters determined from optical SDSS images are affected by dust, especially for objects with small angular size
in the vertical direction \citep{EGIS}. 
Using near-infrared images instead on the optical ones would help mitigate the effects of the dust extinction. 

One more factor that may contribute is that our sample by design consists of 
only thin ($a/b>7$) and bulgeless (Sc--Sd) galaxies.
Our sample with $a/b$ ranging from $7$ to $12$ does not represent all edge-on galaxies with arbitrary disk thickness. 
The effective dynamic range of the disc thicknesses may be not enough to feel the influence on the TF relation.

Note that the relationship between the disk thickness and the spherical-to-disc component mass ratio 
sets only the lower limit on the spherical halo mass \citet{SR2006}. 
In reality many evolutionary factors, such as minor interactions or sources of the internal dynamical heating, deteriorate the relationship. 
Thus, reported small effect of the disk thickness on the TF relation scatter may also reflect contribution of various small-scale factors 
of the disc dynamical heating during galactic evolution. 

We find that the incorporation of the \HI{} and optical colour indexes improves the TF relation scatter 
for edge-on galaxies by about 10 per cent for giant galaxies.
The standard deviation of the mutiparametric TF relation is 0.32 mag in the $r$ and $i$ bands for the galaxies with $\Vmax\ge91$~\kms{}.
It is significantly less than that reported in previous works on the TF relation for flat galaxies.
\citet{KKK1997} constructed the $B_{\rm T}$--$W_{50}$ relation for FGC-galaxies with the standard deviation of 0.58~mag.
In our opinion, a relatively large scatter of TF relation is caused by indirect estimates of the total $B$-band magnitude 
from the angular size and other parameters of edge-on galaxies in the FGC catalogue. 
\citet{KMK+2002} consider statistical properties of the TF relations for flat edge-on galaxies in the $B$, $I$, $J$, $H$ and $K_{S}$ bands.
The near-IR photometry (NIR) was taken from the Two Micron All Sky Survey \citep[2MASS;][]{2MASS}.
They found the r.m.s.{} scatter of 0.58~mag in the $B$-band, 0.48~mag for a deep $I$-photometry and 0.61--0.63~mag in the NIR for a sample of 436 RFGC galaxies.
For studies of large-scale motions in the local Universe, 
\citet{KKK+2003} and \citet{Kashibadze2008} constructed multiparametric TF relations using 2MASS photometry.
\citet{KKK+2003} find the standard deviation $\sigma_{\rm TF}=0.42$~mag for a sample of 971 RFGC galaxies with $\VCMB<18000$~\kms{},
while \citet{Kashibadze2008} obtains the scatter of 0.52~mag for a sample of 410 nearby edge-on galaxies ($\Vh\le3000$~\kms{})
from the 2MASS-selected Flat Galaxy Catalog \citep[2MFGC;][]{2MFGC}.
The quality of these NIR TF relations is obviously limited by short exposures of 2MASS survey,
which leads to an underestimation of the luminosity of galactic discs.
As a rule, accurate and deep photometry in several optical bands significantly improves the TF relation for RFGC galaxies.

The low scatter of our multiparametric TF relation is comparable with the best modern estimations of the TF relation 
for the cases when highly inclined galaxies were excluded from the consideration. 
For example, \citet{SFI++} determine the observational r.m.s.{} scatter of 0.38--0.41~mag for the SFI++ sample of galaxies.
In the frames of Cosmicflows project, \citet{IbandTF} derive the $I$-band TF relation with the r.m.s.{} scatter of 0.41~mag for 267 template galaxies,
and minimize the standard deviation with the 36 zero-point calibrators with Cepheid or TRGB distances down to 0.36~mag.
\citet{Ponomareva+2017} investigate the statistical properties of the TF relation in 12 bands for a sample of 32 galaxies with high quality distances and spatially resolved kinematics. The observational scatter varies from 0.22 to 0.41~mag in the optical and NIR bands.
This literature comparison suggests that the multiparametric TF relation for a RFGC galaxies allows us to measure distances 
with a precision of 0.32~mag, which is not worse than with ordinary samples of galaxies.
We find that the flat edge-on galaxies with extended set of available parameters make 
a good tool to study bulk motions of galaxies in the nearby Universe.

\section*{Acknowledgements}

We thank prof. A.V.\ Zasov for discussions.

The work was carried out under support of the Russian Scientific Foundation (RScF) grant 14--12--00965.
Photometry of galaxies was made within the framework of the grant RScF 14--50--00043.
We acknowledge the usage of the HyperLeda database \citep{LEDA}.


\bibliographystyle{mnras}
\bibliography{tf}   

\appendix{}

\section{The list of the galaxies excluded from our consideration}
\label{appendix:listX}

RFGC\,45 (PGC\,731, UGC\,95) forms a pair with NGC\,27 (UGC\,96) at 1\farcm5.
\HI{}-data are well separated in redshift space, but entangled in different catalogs and databases.
\citet{HGM2011} note that `lower cz feature identified with UGC\,96'.

RFGC\,146 (PGC\,2390, UGC\,418) is a pair with UGC\,422.

RFGC\,268 (PGC\,4278) forms a pair with PGC\,1683298 (2MASX\,J01113021+2312165).

RFGC\,345 (PGC\,5934) is in group with
NGC\,624, $V_h=5871$~\kms{}, at 3\farcm4 and PGC\,5929, $V_h=5916$~\kms{}, at 3\farcm2.

RFGC\,841 (PGC\,15031, UGC\,3027) is a distant giant galaxy at low Galactic latitude.
The EDD spectrum came from \Nancay{} observations \citep{TBCD1998}.
It shows a single peak at 6900~\kms{} with linewidth about 100~\kms{}, which can not fit the morphology of the galaxy.
Observations with Arecibo \citep{PAG1997} show a perfectly two-horned \HI-profile with $W_{50}=461$~\kms.
No doubt, it is the real \HI-spectrum of the galaxy.

RFGC\,1337 (PGC\,23033) is a close pair (angular separation about 30\arcsec) with 2MASX\,J08130371+2433304.

RFGC\,1537 (PGC\,26498, NGC\,2820) forms a group with
NGC\,2814 and NGC\,2805 at distance 3\farcm8 and 13\farcm3 respectively.

RFGC\,1906 (PGC\,32763, NGC\,3454) is a pair with NGC\,3455 at 3\farcm5.

RFGC\,1925 (PGC\,32992, UGC\,6054). 
The foreground star superimposes on the centre of the galaxy and mimics a bright core.
The total magnitude is overestimated.

RFGC\,2130 (PGC\,37054) forms a group with PGC\,37040 at 2\farcm0 and PGC\,37051 at 3\farcm1.

RFGC\,2135 (PGC\,37276, UGC\,6864) is a pair with PGC\,213897 at 2\farcm1.

RFGC\,2329 (PGC\,41974, NGC\,4565C) is a pair with IC\,3546 (PGC\,41976) at 3.8' to S

RFGC\,2817 (PGC\,91372) is a dwarf galaxy with $Vh\approx2200$~\kms{}, confirmed by optical as well as radio observations.
Unfortunately, EDD velocity is 130~\kms{} less then the other data.
The GBT spectrum from EDD is excellent, $W_{50}=455$~\kms{}, but it is too broad for a dwarf galaxy.
There are no giants in wide vicinity of PGC\,91372.
The misidentification or pointing at different object is possible.
\citet{MvD2000} observed the galaxy with \Nancay{} telescope.
Their spectrum has a two-horned profile about 100~\kms{} wide.
This coincides with optical redshift, and the width agrees with the morphology of the galaxy.

RFGC\,4157 (PGC\,91811) is a satellite of UGC\,12714 (RFGC\,4158, PGC\,71969) at 1\farcm8. 
\HI{}-data correspond to UGC\,12714.

\bsp

\label{lastpage}

\end{document}